\documentclass[prl,twocolumn,aps,rsc,longbibliography]{revtex4-1}

\usepackage{amsmath,amssymb,graphicx}
\usepackage{epsf}
\usepackage{latexsym} 
\usepackage{amsmath,amsthm}
\usepackage{ifpdf}
\usepackage{epstopdf}
\usepackage{dcolumn}% Align table columns on decimal point
\usepackage{bm}% bold math
\usepackage{braket}
\usepackage{natbib}

\begin{document}

\def \beq{\begin{equation}}
\def \eeq{\end{equation}}
\def \bea{\begin{eqnarray}}
\def \eea{\end{eqnarray}}
\def \bem{\begin{displaymath}}
\def \eem{\end{displaymath}}
\def \P{\Psi}
\def \Pd{|\Psi(\boldsymbol{r})|}
\def \Pds{|\Psi^{\ast}(\boldsymbol{r})|}
\def \Po{\overline{\Psi}}
\def \bs{\boldsymbol}
\def \bl{\bar{\boldsymbol{l}}}

\title{Quantum simulation of rainbow gravity by nonlocal nonlinearity}
\vspace{0.5cm}

\author{M. C. Braidotti$^{1,2\star}$ and C. Conti$^{2,3}$}
\affiliation{\small
$^1$ Department of Physical and Chemical Sciences, University of L'Aquila, Via Vetoio 10, I-67010 L'Aquila, Italy\\
$^2$ Institute for Complex Systems, National Research Council (ISC-CNR), Via dei Taurini 19, 00185 Rome, Italy\\
$^3$ Department of Physics, University Sapienza, Piazzale Aldo Moro 5, 00185 Rome, Italy\\
$^*$Corresponding author: mariachiara.braidotti@isc.cnr.it}
%\date{\today}

%\begin{abstract}

%\end{abstract}

%\pacs{Nonlinear optics, Quantum simulation}

\maketitle
{\bf %\noindent 150 PAROLE.\\
Testing the unobserved quantum gravitational phenomena in different experimental frameworks is the challenge of analogue gravity. Laboratory emulation may validate theoretical models and give inspiration for further developments. The simulations were limited to general relativity, including black holes, event horizons and superradiance.  We report on the first analog of space- time near a rotating black hole as in a recent quantum-gravity theory, called rainbow gravity. Nonlinear waves in nonlocal media, as those in Bose-condensed gases and nonlinear optics, emulate the rainbow energy-dependent metric.
A fully quantized analysis is reported, showing that the metric energy-dependence inhibits the existence of an event horizon and superradiance. Our results open the way to numerous fascinating experimental tests of quantum gravity theories and demonstrate that these theories can provide novel tools for open problems in nonlinear quantum physics.

} 

%{\bf \noindent Introduction 500 parole.}\\
One of the current major challenges in physics is unifying general relativity and quantum mechanics. Nume\-rous attempts have been carried out, leading to the formulation of many quantum gravity theories\cite{KieferBook,MukhanovBook}, such as string theory\cite{PolyakovBook}, loop quantum gravity\cite{RovelliBook}, non-commutative geometry\cite{ConnesBook} and doubly special relativity\cite{ACameliaDSR_2,MagueijoDSR}. Despite all these efforts, however, it is still not possible to establish which theoretical proposal is the most promising, due to the absence of experimental evidences. As a result, a broad community of scientists is looking for analog systems to provide experimental confirmations of quantum-gravitational phenomena, as Hawking radiation and superradiance\cite{unruh1981,Marino2008_1,Marino2009,Carusotto2008_1,Leonhardt2000,Ornigotti2017,Liberati,FaccioBook,Unruh2005,Unruh2014,Steinhauer2014_1,Fisher2017}. 
This research line has a two-fold advantage: on one hand, emulating these Planck-scale phenomena in the laboratory may suggest unexpected interpretations for quantum gravity, while, on the other hand, analog physics may furnish novel explanations and unexplored regimes for classical and quantum nonlinear physics.\\
The first analog dates back to 1981 \cite{unruh1981} when Unruh proposed black-hole evaporation as a model to study sound waves in moving fluids, showing that it is possible to find Hawking radiation in a non-gravitational system. This remarkable finding opened the way to a wide investigation of analogue gravitational phenomena.\\
Many authors after Unruh reported emulations of black holes in many research fields as acoustics\cite{unruh1981}, optics\cite{Unruh2005,Leonhardt2000,Philbin2008}, Bose-Einstein condensates \cite{Garay_1,Barcelo2003,Giovanazzi2004}, $^3He$ \cite{Volovik} and Fermi liquids\cite{Giovanazzi}, as reviewed in \cite{Liberati}. 
However, all these studies mainly address to phenomenological aspects, not providing frameworks to test the numerous formulations of quantum gravity. To our knowledge, an analogue of a quantum gravity scenario has never been proposed.\\
In this study, we report on classical and quantum analogs of a recent theory of quantum gravity called ``rainbow gravity''. This theory is a generalization of doubly special relativity\cite{ACameliaDSR_2} to incorporate curvature. Rainbow gra\-vity, proposed by Magueijo and Smolin in 2004\cite{magueijo2002_2,Magueijo_rainbow}, relies on a space-time geometry, which depends on the energy of the free-falling particle. 
This corresponds to an energy-dependent metric with invariant line element:
\begin{equation}
ds^2=g_{\mu\nu}(E)dx^{\mu}dx^{\nu}=-\frac{(dx^0)^2}{f^2(E/E_P)}+\frac{(dx^i)^2}{g^2(E/E_P)},
\label{invariant-line}
\end{equation}
with $E_P$ the Planck energy.
The functions $f$ and $g$ enclose all the metric energy-dependence and their form affects the space-time properties, as inducing a non-constant light speed $c$ and/or an energy-dependent gravi\-ta\-tio\-nal constant.
Some of the implications of Eq.(\ref{invariant-line}), as in black hole thermodynamics, are not known and inven\-ting emulations may furnish many unexpected results and be an inspiration for further developments.\\
In this paper, we study the behavior of excitations on a vortex background in a defocusing nonlinear nonlocal medium. We show that nonlocality allows to mimic the rainbow gravity space-time of a rotating black hole, where the degree of nonlocality $\sigma$ determines the pro\-xi\-mity to the Planck scale. Numerical simulations of classical and quantum excitations give evidence of the fading of the black hole event horizon and the consequent wea\-ke\-ning of superradiance when increasing $\sigma$. Full quantum dynamics is analyzed in the $\mathcal{P}$-representation by a pseudo-spectral stochastic Runge-Kutta algorithm\cite{DrummondBook} and shows an enhancement of the vanishing of the ergoregion. This furnishes a true quantum-simulation of second-quantized fields in a curved space-time with ener\-gy dependent metric.\\

%%%%%%%%%%%%%%%%%%%%%%%%%%%%%%%%%%%%%%%%%%%%%%%%%%%%%% \section{THEORY}
%{\bf \noindent Main 2000$/$3000 parole.}\\
We start considering the classical regime and the way an energy dependent metric occurs in the hydro\-dy\-na\-mi\-cal approximation of the normalized nonlocal nonlinear Schr\"odinger equation $[\bm{r}=(x,y)]$ \cite{Braidotti_GUP16,Belic2012,Krolikowski04}
\begin{equation}
\imath\partial_t \psi+\frac{1}{2}\nabla_{xy}^2 \psi -PR(\bm{r})\ast|\psi|^2\psi=0.
\label{nls_ad}
\end{equation}
Equation (\ref{nls_ad}) describes the field evolution in many phy\-si\-cal systems as nonlinear optics with thermal\cite{Gentilini:12_1,Braidotti_GUP16} or re-orientational nonlinearity, Bose-Einstein condensates (BEC)\cite{Hoefer2006,Dominici2015} and plasma-physics\cite{Taylor1970,Romagnani2008}. In (\ref{nls_ad}), $\ast$ denotes a convolution integral. The form of the kernel $R(x,y)$ depends on the specific physical system and its Fourier transform is $\tilde R(K_x,K_y)$. 
The field $\psi$ is norma\-lized such that $\int |\psi|^2d\bm{r}=1$, and $P$ measures the strength of the nonlinearity.\\
Thanks to the hydrodynamical approach, commonly used for dispersive shock waves \cite{Gentilini:12_1} and also analog gravity \cite{Fouxon2010}, we study the behavior of small excitations on top of the metric induced by the fluid of light. By wri\-ting the field $\psi$ as $\psi = \sqrt{\rho}e^{\imath\phi}$ with $\bm{v}=\nabla\phi$, Eq. (\ref{nls_ad}) reduces to the continuity equation and the Euler equation,
with bulk pressure $\mathcal{P}[\rho] = \rho P R(\bm{r})\ast\rho$ (see Supplementary Information). 
Small excitations in the photon fluid are described by letting
$\rho = \rho_0 +\epsilon\rho_1 + O(\epsilon^2)$ and $\phi= \phi_0 +\epsilon\phi_1	+ O(\epsilon^2)$. 
In a slowly varying background $\rho_0$ and in the eikonal approximation, we have $\rho_1=\bar{\rho_1}e^{\imath(K_x x+K_y y-E t)}$ and $\phi_1=\bar{\phi_1}e^{\imath(K_x x+K_y y-E t)}$, with $E$ the angular frequency for an inertial observer at infi\-ni\-ty.
The generalized dispersion relation is (see Supplementary Information) \cite{Bogoliubov47,Dalfovo99,Picozzi2014,Chiao1999}:
\begin{equation}
(E-\bm{K}\cdot \bm{v_0})^2=\rho_0P\tilde{R}(K) K^2+\frac{K^4}{4}.
\label{bog}
\end{equation} 
In Eq. (\ref{bog}) $\bm{v}_0=\nabla \phi_0$ is the background velocity field.
When $\bm{v_0}=0$, the high energy limit, i.e., short wavelengths and large momentum $K=\left(K_x^2+K_y^2\right)^{1/2}$, corresponds to $E\propto K^2$ as for free particles. Whereas, in the long wavelength limit (i.e. for small $K$) $E\simeq c_s K$, where $c_s=P\rho_0$ is the local speed of sound $c_s^2\equiv\left.\frac{\partial\mathcal{P}[\rho]}{\partial\rho}\right\vert_{\rho=\rho_0}$.\\     
In the hydrodynamical regime, we obtain the following equation for the Fourier transformed massless scalar field $\tilde \phi_1$ in a $2+1$ dimensional curved space (see Supplementary Information):
\begin{equation}
\widetilde{\Delta\phi_1}=\frac{1}{\sqrt{-\tilde{g}}}\partial_{\mu}(\sqrt{-\tilde{g}}\tilde{g}^{\mu\nu}\widetilde{\partial_{\nu}\phi_1})\text{,}
\label{kge}
\end{equation}
where $\tilde{g}^{\mu\nu}$ is the covariant metric $\left[(\tilde{g}^{\mu\nu})^{-1}=\tilde{g}_{\mu\nu}\right]$,
\begin{equation}
\tilde{g}_{\mu\nu}=\left(\frac{\tilde{R}}{c_s^2}\right)\left(\begin{matrix}
-(c_s^2\tilde{R}-v_0^2) & -\bm{v_0^T}\\
-\bm{v_0} & I
\end{matrix}\right)
\label{metric}
\end{equation}
and $\tilde{g} = \mbox{det}(\tilde{g}_{\mu\nu})$. In (\ref{metric}), $I$ is the $2\times2$ identity matrix. Equation (\ref{kge}) gives the analogy between the light wave propagation and the gravitational field: the light fluc\-tua\-tions behavior is affected by the metric $g_{\mu\nu}$ induced by the background $\rho_0$. \\
The geometrical description of the optical system fails when the background density varies on a scale smaller than the healing length $\xi=1/(2\sqrt{P\tilde{R}\rho_0})$, i.e. in the high energy limit where $E\propto K^2$. In terms of the quantum-gravity analog, the scale of background density variation exceeds the analog Planck length $\xi$, violating the Lorentz invariance. Hence, the analogue is self-consistent only in the low energy and low-momentum regime for which $K<1/\xi$.\\
Following general relativity, a metric determines the invariant square of a line element, $ds^2$, given by 
\begin{align}
&ds^2=\tilde{g}_{\mu\nu}dx^{\mu}dx^{\nu}=\nonumber\\
&=\left(\frac{\tilde{R}}{c_s^2}\right)\Biggl\{-\left(c_s^2\tilde{R}-v_0^2\right)dt^2 +\Biggr.\label{ds}\\
&+ \Biggl.\Biggl[dr^2+(r d\theta)^2\Biggr]-2v_rdrdt-2v_{\theta}rd\theta dt \Biggr\}\nonumber.
\end{align}
Equation~(\ref{ds}) is written in polar coordinates ($r,\theta$), with $v_0^2= v^2_r + v^2_{\theta}$, $v_r = \partial_r\phi_0$ and $v_{\theta}= \frac{1}{r}\partial_{\theta}\phi_0$. The local case \cite{Marino2008_1,*Marino2009} is found letting $\tilde R=1$ in Eq. (\ref{metric}).\\ 
Within the geometrical description validity limit region ($K<1/\xi$), we can mimic a black hole through vortex solutions of Eq.~({\ref{nls_ad}}). This, in addition with the presence of nonlocality, enables to simulate rainbow gravity.\\
Vortexes are formed by a dark hole with circular symmetry and a helical wave front. Their  wavefunctions, in polar coordinates, is $\psi_0=\sqrt{\rho_0(r)}e^{i\phi_0(\theta)}$, with phase $\phi_0=m\theta$. The integer $m$ is the winding number, or vortex charge. The region in proximity of $r=0$, where $\rho(0)=0$, is called the vortex core.\\ 
The form of the metric $\tilde g_{\mu, \nu}$ determines the properties of the analog space-time, as the presence of an event horizon and ergoregion.\\
The analog system presents an event horizon surface if the velocity of sound equals the radial velocity of the fluid, i.e. $c_s=v_r$. Low energy modes cannot escape from this region. Furthermore, by geometrical considerations, one can find that $\tilde g_{tt}\geq0$, and hence  $c_s^2<v_0^2$, determines the ergoregion, which is the region of space-time where low energy modes are dragged in the moving light flow.\\
In the case of local response function $\tilde R=1$, this geo\-me\-try has an ergosurface placed at $c_s=v_{\theta}$. In order to mimic a black hole, it is necessary to introduce a radial inward velocity $v_r$, by letting $\phi_0(r,\theta)=2\pi\sqrt{r/r_0}+m\theta$, which induces an event horizon \cite{Marino2008_1,*Marino2009,visser2000comment,Brevik2001}. In this frame, the position of the event horizon $r_H$ is $\pi^2/r_0 c_s^2$ and the ergoregion $r_E=\frac{1}{2}\left(r_H+\sqrt{r_H^2+4m^2r_0^2r_H^2}\right)$.
Unfortunately, such vortex solution with a central sink in a nonlinear medium with local response function is not stable \cite{Skryabin1997,Kivshar2005} and hence it does not allow experimental tests of the analogy.
Whereas, in the nonlocal case, the stability of the vortex solitons has been reported \cite{Assanto2015}. Nonlo\-ca\-lity opens the way to many theoretical and experimental developments.\\
In order to provide an analog of a rotating black hole and analyze the effects of a rainbow gravity space-time, we study the propagation of a pulse with group velocity $v_g=\partial_K\omega$ in a nonlocal medium. We remark that nonlocality provides the stability of the vortex background, allowing mimicking the Kerr-type black hole metric, while, the different frequency components of the pulse spectrum will help in testing the energy space-time dependence of rainbow gravity.\\
For a narrow-band wavepacket with mean momentum $K$ in a nonlocal medium and in the small $K$ limit, the energy $E$ is proportional to $K$ and the metric in Eq.~(\ref{ds}) can be written in terms of the energy as in the original rainbow gravity theory. (In our units $E_P=1$.)
We show in the following that our system emulates a rainbow gravity theory and the metric can be written as the corresponding Kerr black hole metric\cite{Magueijo_rainbow,magueijo2002_1,*magueijo2002_2,Zhao2016}, i.e.,
\begin{equation}
ds_r=-\frac{T(r)}{f^2(E)}dt^2+\frac{R(r)}{g^2(E)}dr^2+\frac{r^2d\theta^2}{g(E)^2}+V_{\theta}(r)\frac{rd\theta dt}{f(E)g(E )}
\end{equation}
where the functions $T(r)$, $R(r)$ and $V_{\theta}(r)$ are the elements of the energy-independent metric matrix. 
By rescaling time and azimuthal coordinates far from the vortex core, we find that, in our case,
%Far from the vortex core, we rescale the time and azimuthal angle as
%\begin{eqnarray}
%dr=d\varrho\mbox{;}&\qquad& dt=d\tau-\frac{f}{g}\frac{v_r}{(c_s^2-v_r^2)}d\varrho\\
%r d\theta&=&\varrho d\vartheta-\frac{v_r v_{\theta}}{(c_s^2-v_r^2)}d\varrho
%\end{eqnarray}
%and we write Eq. (\ref{ds}) as
%\begin{eqnarray}
%ds^2&=&-\left(c_s^2-v_0^2\right)\frac{d\tau^2}{f^2(E)} +\\
%&-& \frac{c_s^2}{c_s^2-v_r^2}\frac{d\varrho^2}{g^2(E)} +\frac{(\varrho d\vartheta)^2}{g^2(E)}-2v_{\theta}\frac{\varrho d\vartheta d\tau}{f(E)g(E)},
%\end{eqnarray}
%where 
the functions $f(E)$ and $g(E)$ are
\begin{equation}
f(E)=\frac{c_s^2}{\tilde{R}}\frac{c_s^2-v_0^2}{c_s^2\tilde{R}-v_0^2}\mbox{;} \qquad g(E)=\frac{c_s^2}{\tilde{R}}\frac{c_s^2\tilde{R}-v_0^2}{c_s^2-v_0^2}. \label{fg}
\end{equation}
Details on the calculus are reported in the Supplementary Information.
Equation~(\ref{fg}) shows that, in case of nonlocal response, the position of the event horizon and ergoregion are energy independent, while the horizon area is energy-dependent, i.e., in the 2D case we have
\begin{align}
\mathcal{A}&=\int_0^{2\pi}d\theta \sqrt{g}\bigg|_{r=r_H}=\int_0^{2\pi}d\theta \sqrt{\frac{r}{g(E)}}\bigg|_{r=r_H}=\\
&=\frac{2\pi r_H v_{\theta}}{c_s}\sqrt{\frac{\tilde{R}}{c_s^2\tilde{R}-v_0^2}}
\label{HArea}
\end{align}
since at $r=r_H$ we have that $c_s=|v_r|$. The dependence of $\mathcal{A}$ on the energy $E$ is due to nonlocality. This is in agreement with a key-prediction of rainbow gravity\cite{Magueijo_rainbow}.\\
Indeed, in the nonlocal case, particles with different e\-ner\-gy see different horizon areas: Fig. 1a shows a sketch of the fading of the area of a rotating black hole near the Planck energy scale in rainbow gravity. In the standard local case $(\tilde R=1)$, there is no energy dependence.\\
In order to describe different physical systems as optical nonlinear waves and Bose-Einstein condensates, we consider a nonlocal response function of the form
\begin{equation}
\tilde{R}(K)=B+\frac{1}{1+\sigma^2K^2}, 
\label{response}
\end{equation}
with $\sigma$ the degree of nonlocality.
%%%%%%%%%%%%%%%%%%%%%%%%%%%%%%% FIGURE 1 %%%%%%%%%%%%%%%%%%%%%%%%%%%%%%%%%%%%%
\begin{figure*}[t!]
	\centering
		\includegraphics[scale=0.6]{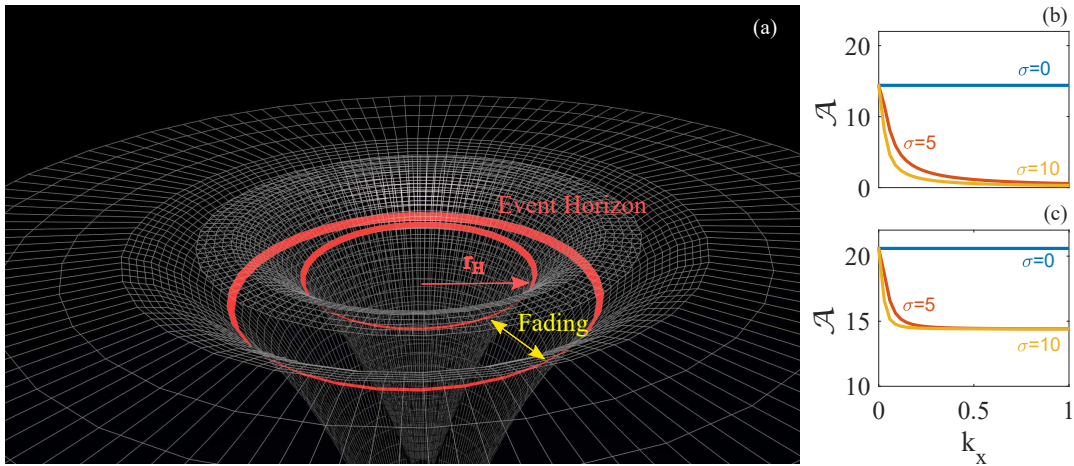}
	\centering
	\caption{(Color online) (a) Sketch of the black hole in rainbow gravity. Red lines correspond to event horizons in the local ($\sigma=0$-inner horizon) and nonlocal cases ($\sigma\neq 0$-outer horizon) for a specific $K$; (b) horizon area $\mathcal{A}$ as function of $K_x$ for lorentzian nonlocal response function $(B=0)$ varying $\sigma$; (c) as in (b) for BEC-like response function $B=1$.}
\label{fig:1}
\end{figure*}
%%%%%%%%%%%%%%%%%%%%%%%%%%%%%%%%%%%%%%%%%%%%%%%%%%%%%%%%%%%%%%%%%%%%%%%%%%%%%%%%%%%%%%%%
In (\ref{response}), $B=0$  corresponds to a Lorentzian spectral response characteristic of optical media with thermal or re-orientational nonli\-nea\-rities; while, in the case $B=1$, the nonlocal response is composed by a local and a nonlocal contribution, 
as occurs in photonic BEC. \cite{Calvanese2014,weitz2010}
Figure \ref{fig:1}b shows the horizon area $\mathcal{A}$ trend for Lorentzian response as function of the transverse wavevector $K_x$ for different values of $\sigma$. We observe that nonlocality affects the value of the horizon area and hence, wavepackets with different momentum see different areas. In the following, we will see that this kind of behavior resembles what happens in a rainbow gravity scenario. In fig. \ref{fig:1}b the horizon area $\mathcal{A}$ decreases with $K_x$ in the nonlocal case. 
In Fig. \ref{fig:1}c, when $\sigma\neq0$ the horizon area $\mathcal{A}$ saturates at a lower value when $K_x$ growths, i.e. the horizon area fades into a volume. In the following, we show that these findings affect the black hole thermodynamics. 

\noindent In this frame, a notable effect that may arise is super\-ra\-dian\-ce, which takes place in proximity of the event horizon of a rotating black hole. Superradiance is the amplification of radiation excitations due to the angular velocity of the vortex. \cite{Ornigotti2017} Being rainbow gravity a recent theo\-ry, there is not a comprehensive vision about its phenomenology, hence the occurrence of superradiance has still not been fully addressed. Despite this, in the rainbow gravity scenario, we expect that superradiance is reduced because of the energy dependent coupling and the fading of the event horizon area. We are not aware of previously reported analysis of analog rainbow superradiance.\\ 
In order to analyze superradiance, we consider a perturbation $\phi_1$ of the form $\phi_1(z,r,\theta)=r^{-1/2}G(r^*)e^{i(\Omega t-n\theta)}$ such that it is solution to the Klein-Gordon equation (\ref{kge}) with metric (\ref{metric}), where $n$ is the winding number and $\Omega$ the wave frequency. It is worth to change the coordinate system, adopting the ''tortoise coordinate'' $r^*$, which maps the region $r\in[r_H,\infty[$ to the entire axis. Tortoise coordinates are defined as $dr^*=(1-r_H/r)^{-1}dr$.  Note that $r^*$ is defined only for $r>r_H$. As $r$ approaches the event horizon $r_H$, $r^*\rightarrow-\infty$, while far from the vortex core $(r\rightarrow\infty)$ we have $r^*\rightarrow r$. After some algebra (see Supplementary Information) we find the Schr\"odinger-like equation for the radial component $G(r^*)$ 
\begin{equation}
\partial_{r^*}^2 G+V_{eff} G=0,
\label{schr}
\end{equation}
where the effective potential $V_{eff}$ is given by
\begin{align}
&V_{eff}=\left(\frac{\Omega}{c_s\sqrt{\tilde R}}-\frac{v_{\theta}n}{r c_s\sqrt{\tilde R}}\right)^2+\nonumber\\
&-\frac{dr}{dr^*}\left(\frac{1}{2}\frac{r_H}{\tilde{R}r^3}-\frac{n^2}{r^2}\right)+\left(\frac{dr}{dr^{*}}\right)^2\frac{1}{4r^2},
\label{Veff}
\end{align}
where we assumed $\partial_r \tilde{R}\simeq 0$.\\
In order to compute superradiance, we analyze Eq. (\ref{schr}) in two limits $r^*\rightarrow\pm\infty$ and find that the effective potential is
\begin{align}
&\mbox{for }r\rightarrow\infty\qquad V_{eff}=\left(\frac{\Omega}{c_s\sqrt{\tilde R}}\right)^2,\\
&\mbox{for }r\rightarrow r_H \qquad V_{eff}=\left(\frac{\Omega}{c_s\sqrt{\tilde R}}-\frac{v_{\theta}n}{r_H c_s\sqrt{\tilde R}}\right)^2,
\label{Veff_lim}
\end{align}
and hence we have
\begin{align}
&\mbox{for }r\rightarrow\infty\qquad G(r^*)=e^{i\frac{\Omega}{\sqrt{\tilde{R}_{\infty}}c_s} r^*}+\mathcal{R}e^{-i\frac{\Omega}{\sqrt{\tilde{R}_{\infty}}c_s} r^*}\label{G_lim1},\\
&\mbox{for }r\rightarrow r_H \qquad G(r^*)=\mathcal{T}e^{i\frac{(\Omega-n\Omega_H)}{\sqrt{\tilde{R}_H}c_s}r^*},
\label{G_lim2}
\end{align}
where $\Omega_H=\frac{v_{\theta}}{r_H}$ is the angular velocity at the horizon and $\mathcal{R}$ and $\mathcal{T}$ are the reflection and transmission coefficients.  
Eq. (\ref{G_lim2}) accounts only for the ingoing wave at the horizon since the outgoing mode can not be considered as a physical solution. 
By the Abel's theorem, the Wronskian of Eqs. (\ref{G_lim1}) and (\ref{G_lim2}) is constant. Hence, equating the Wronskian computed at the two limits, we get the relation between $\mathcal{R}$ and $\mathcal{T}$:
\begin{equation}
1-|\mathcal{R}|^2=\frac{\Omega-n\Omega_H}{\Omega}\sqrt{\frac{\tilde{R}_{\infty}}{\tilde{R}_H}}|\mathcal{T}|^2.
\label{sr}
\end{equation}
Equation (\ref{sr}) shows that, if the frequency $\Omega$ of the incident perturbation is in the range $0<\Omega<n\Omega_H$ and if $\frac{\tilde{R}_{\infty}}{\tilde{R}_H}>0$, the amplitude of the scattered wave is larger than that of the incident one, i.e., $|\mathcal{R}|>1$. Hence the perturbation is superradiantly amplified in analogy with superradiant scattering from a rotating black hole. \\

%%%%%%%%%%%%%%%%%%%%%%%%%%%%%%%%%%%%%%%%%%%%%%%%%%%% 

\noindent {\bf Classical simulations}. As previously said, rainbow gravity is a recent theory of quantum gravity not yet fully developed. Because of this, we resort to numerical simulations in order to suggest specific experimental directions. 
%%%%%%%%%%%%%%%%%%%%%%%%%%%%%%% FIGURE 2 %%%%%%%%%%%%%%%%%%%%%%%%%%%%%%%%%%%%%
\begin{figure*}[t!]
	\centering
		\includegraphics[scale=0.8]{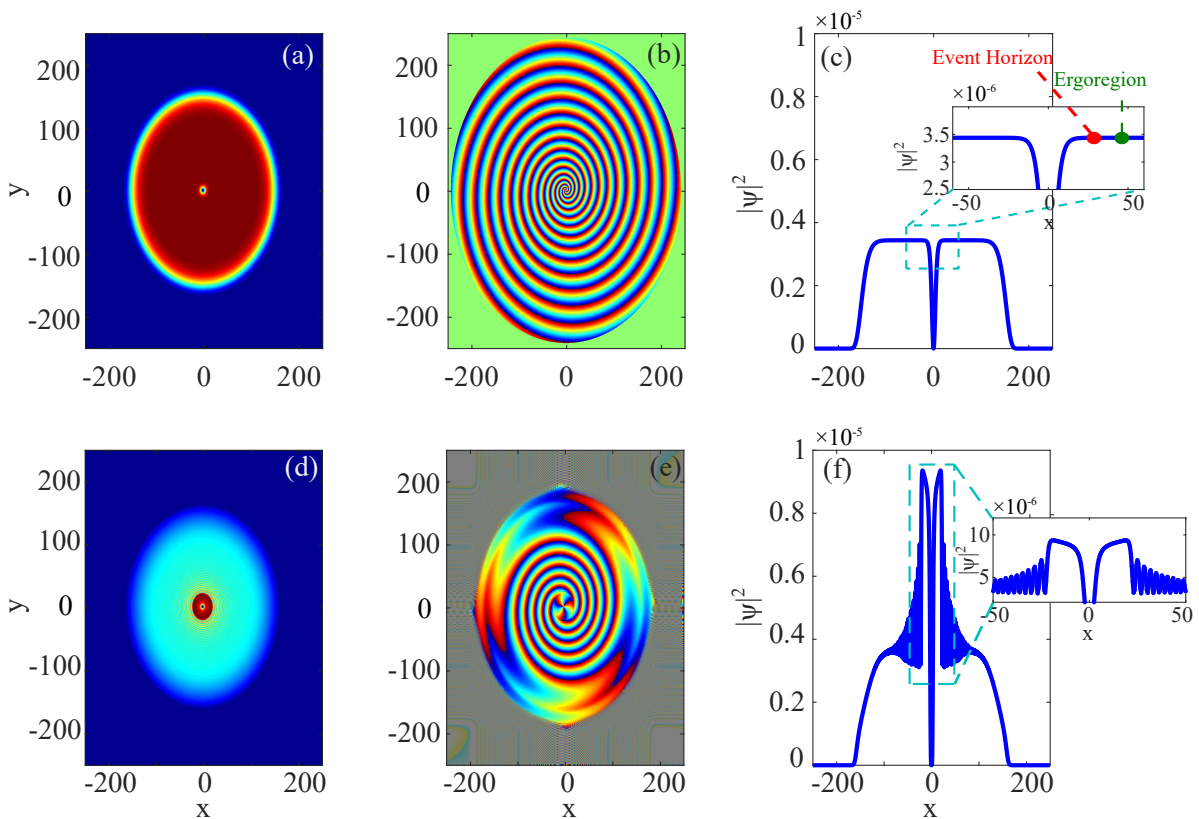}
	\centering
	\caption{(Color online) (a) intensity field profile for $t=0$; (b) phase profile for $t=0$; (c) field section along $x$ for $y=0$ for $t=0$; inset: corresponding enlarged central region of the vortex profile at $t=0$, dots mark the location of the event horizon $(x\simeq28$ - red) and ergoregion $(x\simeq46$ - green);  (d) as in (a) for $t=30$; (e) as in (b) for $t=30$; (f) as in (c) at $t=30$; inset: corresponding enlarged central region of the vortex profile at $t=30$. (The beam parameters are $w_0=160$, $w_v=5$, $m=3$, $\sigma=0$ and $P=10^5$)}
\label{fig:2}
\end{figure*}
%%%%%%%%%%%%%%%%%%%%%%%%%%%%%%%%%%%%%%%%%%%%%%%%%%%%%%%%%%%%%%%%%%%%%%%%%%%%%%%%%%%%%%%%
To our knowledge, simulations of analog superradiance have not been addressed yet.\\ 
In the following, we first analyze classical superradiance in the local and nonlocal case, then we will consider the fully-quantum counterpart.\\
We simulate the propagation of a vortex beam in a defocusing nonlinear nonlocal medium by the classical nonlinear Schr\"odinger equation, Eq. (\ref{nls_ad}) (see Methods). The initial condition is
$\psi_0=N \exp[-(r/w_0)^{16}] \tanh\left(\frac{r}{w_{v}}\right)e^{i\phi_0}$
with $w_v$ is the vortex waist and $N$ is a normalization constant. $\psi_0$ includes a finite supergaussian background with waist $w_0\gg w_v$. The vortex velocity  $\bm v=(v_r,v_{\theta})$ is composed by the radial component $v_r=\pi/\sqrt{r r_0}$, which is linked to the event horizon, and the azimuthal component $v_{\theta}=m/r$. \\
The considered initial condition $\psi_0$ is not an exact solution of Eq. (\ref{nls_ad}). For this reason, it can be written as the exact solution $\bar\psi$ of Eq. (\ref{nls_ad}) plus an additional term $\psi'$: $\psi_0=\bar\psi+\psi'$.
This fact, in the presence of the radial velocity field, makes our configuration an optimal analog for superradiance. 
Figure \ref{fig:2} shows the intensity spot profile, its phase and the $x$ section of $|\psi|^2$ at $t=0$ (up line panels) and at $t=30$ (down line panels). Inset of panel \ref{fig:2}c shows the positions of the event horizon $(x\simeq28)$ and ergoregion $(x\simeq46)$ of our black hole analog. Comparing panels \ref{fig:2}c and \ref{fig:2}f, we see that during the beam evolution the intensity near the vortex core increases and the emitted radiation profile exhibits field oscillations. Furthermore, as expected for superradiance, the oscillations' amplitude decreases moving away from analog black hole.\\
We then consider the effect of the energy dependent metric on the Bogoliubov dispersion relation for local and nonlocal nonlinearities.
%%%%%%%%%%%%%%%%%%%%%%%%%%%%%%% FIGURE 3 %%%%%%%%%%%%%%%%%%%%%%%%%%%%%%%%%%%%%
\begin{figure*}[t!]
	\centering
		\includegraphics[scale=0.8]{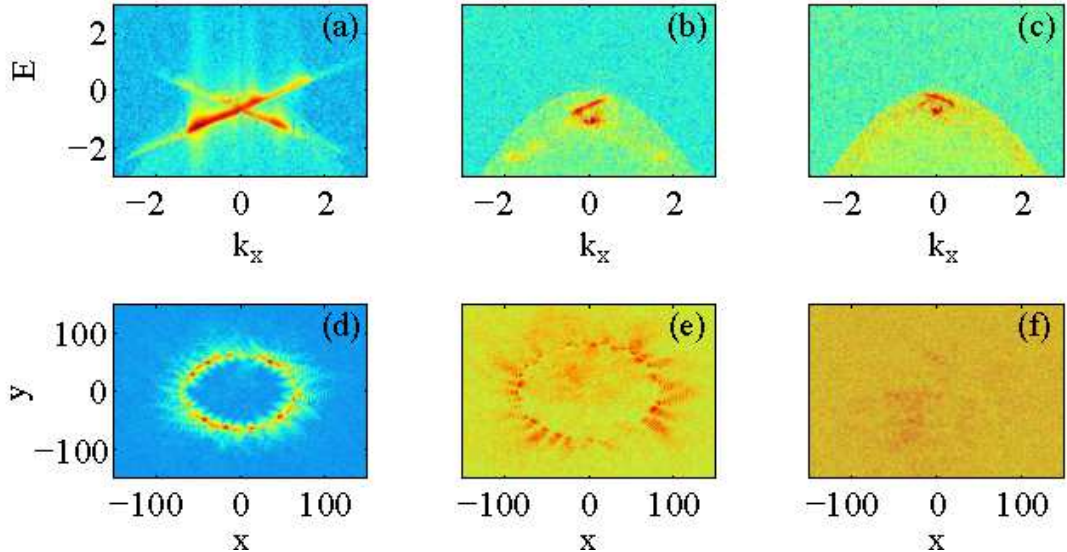}
	\centering
	\caption{(Color online) (a) excitations spectrum in the local case; (b) as in (a) in the nonlocal case ($A=1$ and $\sigma=5$); (c) as in (b) for $A=0$ and $\sigma=5$; (d) excitations field in the configuration space in the local case; (e) as in (d) for $A=1$ and $\sigma=5$; (f) as in (e) for $A=0$ and $\sigma=5$. ($P=10^5$)}
\label{fig:3}
\end{figure*}
%%%%%%%%%%%%%%%%%%%%%%%%%%%%%%%%%%%%%%%%%%%%%%%%%%%%%%%%%%%%%%%%%%%%%%%%%%%%%%%%%%%%%%%%
We add to the initial $\psi_0$ a classical noise, which mimics fluid excitations. Figure \ref{fig:3} shows the excitations spectrum and the event horizon at $t=30$ in the local and nonlocal cases. Different nonlocal response functions are taken into account. For local nonlinearity $(\tilde R=1)$, the dispersion exhibit a linear trend as expected for particles trapped in the event horizon (Fig. \ref{fig:3}a). This corresponds to a preferred space location in the $(x,y)$ space (circle in Fig. \ref{fig:3}d). In the nonlocal case ($A=1$), Figs. \ref{fig:3}b and \ref{fig:3}e show the fading of the event horizon and the destruction of the linear Bogoliubov spectrum. Figures \ref{fig:3}c and \ref{fig:3}f show the case $A=0$ with the disappearance of the event horizon: the perturbation behaves as free particles homogeneously distributed in space (Fig. \ref{fig:3}f). \\
%%%%%%%%%%%%%%%%%%%%%%%%%%%%%%%%%%%%%%%%%%%%%%%%%%%%5

\noindent {\bf Quantum simulations}. So far, we have considered only classical analogs of the black hole. It is extremely relevant to show the possibility of a true quantum simulation of the analog black hole in order to test the rainbow gravity in a true quantum scenario. To this aim, we resort to the second quantized nonlocal nonlinear Schr\"odinger equation
\begin{equation}
\imath\partial_t \hat\psi+\frac{1}{2}\nabla_{xy}^2 \hat\psi -P\left[R_1 \star \hat\psi^\dagger\hat\psi\right]\hat\psi=0.
\label{qnls_ad}
\end{equation}
The quantum field $\hat\psi$ obeys the equal time commutation operation $\left[\hat\psi(\bm{r},t),\hat\psi\dagger(\bm{r}',t)\right]=\delta(\bm{r}-\bm{r}')$.
Quantum fields are mathematically described by operator di\-stri\-butions, and hence the equations which govern their evolutions are operator equations that can not be numerically solved. However, operators distributions can be expressed by phase-space representations that map operator equations to equivalent stochastic differential equations. We adopt the positive $\mathcal{P}$-representation that transforms the Heisenberg (operator) equations of motion in a Fokker-Planck equation (FPE).\cite{DrummondBook} The positivity of the representation allows to map the FPE to It\^o stochastic differential equations (details are reported in Methods):
\begin{eqnarray}
\frac{\partial}{\partial z}u&=&+i\frac{1}{2}\nabla^2_{xy}u-iP\left[R(r)\ast uv\right] u+\sqrt{i}\Gamma^{(1)}u \label{ito_u}\\
\frac{\partial}{\partial z}v&=&-i\frac{1}{2}\nabla^2_{xy}v+iP\left[R(r)\ast uv\right] u+\sqrt{-i}\Gamma^{(2)}v \label{ito_v}
\end{eqnarray} 
where $\Gamma^{(i)}(r,z)$ is a real Gaussian white noise.
We solve (\ref{ito_u}) and (\ref{ito_v}) by a second-order pseudo-spectral stochastic Runge-Kutta algorithm.\cite{DrummondBook} \\
%%%%%%%%%%%%%%%%%%%%%%%%%%%%% FIGURE 5 %%%%%%%%%%%%%%%%%%%%%%%%%%%%%%%%%%%%%%%%%%%%%%%%%%%
\begin{figure*}[t!]
	\centering
		\includegraphics[scale=0.8]{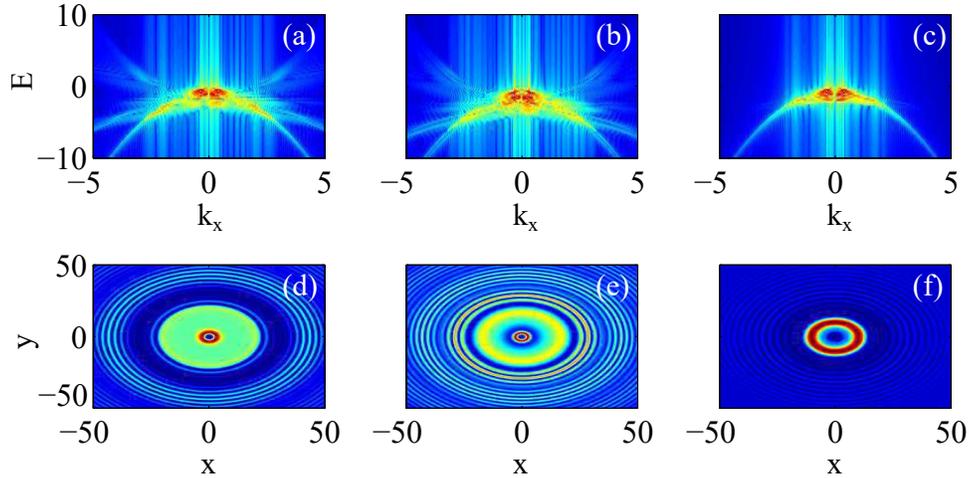}
	\centering
	\caption{(Color online) (a) quantum spectrum in the local case; (b) as in (a) in the nonlocal case ($A=1$ and $\sigma=5$); (c) as in (b) for $A=0$ and $\sigma=5$; (d) quantum field in the configuration space in the local case; (e) as in (d) for $A=1$ and $\sigma=5$; (f) as in (e) for $A=0$ and $\sigma=5$.}
\label{fig:5}
\end{figure*}
%%%%%%%%%%%%%%%%%%%%%%%%%%%%%%%%%%%%%%%%%%%%%%%%%%%%%%%%%%%%%%%%%%%%%%%%%%%%%%%%%%%%%%%%%%
Figure \ref{fig:5} shows the excitation field spectra and intensities of local and nonlocal case for $t=30$. Comparing the panels \ref{fig:5}a, \ref{fig:5}b and \ref{fig:5}c, we see that Figs. \ref{fig:5}a and \ref{fig:5}b exhibit linear sidebands which are not present in the nonlocal lorentzian spectrum (Fig. \ref{fig:5}c). These sidebands can be attributed to the quantum radiation in proximity to the event horizon.
Superradiance is also present. Panel \ref{fig:5}d shows the presence of an ergoregion in the local frame. This region fades progressively with the increasing of the nonlinear effect (see panels \ref{fig:5}e and \ref{fig:5}f). Furthermore the central region of the spectrum in Fig. \ref{fig:5}a exhibits a linear dispersion that can be attributed to the radiation in proximity of the event horizon. A key difference between the classical and the quantum analysis, is that, in the latter case, the spectral content of the noise is much wider, because the quantum noise is continuously generated upon evolution. This is evident in Fig. \ref{fig:5}a, with respect to the classical case in Fig. \ref{fig:3}a, but the key point is that the spectrum of the trapped quantum excitation fades and then completely disappears in the rainbow gravity case. In addition, in the quantum case we see the vanishing of the ergoregion not evident in the classical case (see Figs. \ref{fig:3}e and \ref{fig:5}e). \\
%%%%%%%%%%%%%%%%%%%%%%%%%%%%%%% \section{CONCLUSIONS} 
In conclusion, we propose the first analog of rainbow quantum gravity by nonlocal nonlinear waves. We theoretically describe a black hole as a stable nonlocal vortex and show the fading of the event horizon and the inhibition of superradiance in classical and second quantized frameworks. Our findings can trigger further research in quantum gravity by novel experimental and theoretical emulations. Since several open questions are still present about rainbow gravity and competing theories, we think that quantum simulations may provide new surprising insights.\\

%\vspace*{0.2cm}

\noindent {\bf  Methods}
%\vspace*{0.2cm}
\small{
Classical simulations: we simulate the NLS equation through the split-step Fourier method with noisy initial condition. In order to calculate the Bogoliubov dispersion relation we subtract the evolved noisy field to the unperturbed solution at the same instant of propagation $t$, obtaining the evolved noise. We mediate over several (n=20) noise configurations.\\
Quntum simulations: The system has been simulated with the stochastic Runge-Kutta algorithm \cite{DrummondBook}: second order in the deterministic part and order 1.5 in the stochastic part with 20 disorder averages.
The derivatives in the deter\-mi\-ni\-stic part are computed using the fast Fourier transform (FFT) algorithm; the second derivatives of the field with respect to the variable $x$ is computed by inverse Fourier transform.
The noise realizations have the form $dW_k \approx N(0,N_{x})\sqrt{\Delta z/\Delta x}$ , where $N_x$ is the number of points in the discretization of the variable $x$ and $N$ is a random number normally distributed between $0$ and $N_x$.
}

%%%%%%%%%%%%%%%%%%%%%%% References %%%%%%%%%%%%%%%%%%%%%%%%%
%\bibliography{references}% Produces the bibliography via BibTeX.
%\bibliographystyle{rsc}
%\bibliographystyle{unsrt} 
%\bibliographystyle{apsrev} 

%\end{thebibliography}%
%merlin.mbs apsrev4-1.bst 2010-07-25 4.21a (PWD, AO, DPC) hacked
%Control: key (0)
%Control: author (0) dotless jnrlst
%Control: editor formatted (1) identically to author
%Control: production of article title (0) allowed
%Control: page (1) range
%Control: year (0) verbatim
%Control: production of eprint (0) enabled
%

\noindent {\bf Acknowledgments}\\ 
\small{We acknowledge support from the John Templeton Foundation (grant 58277).\\}

\noindent {\bf Author contribution}\\
\small{All authors conceived the idea. M.C.B. carried out the numerical simulations and data analysis with contributions from C.C.. All authors developed the interpretation of results and discussed the results. All authors  wrote the paper.\\}

\noindent {\bf Supplementary Information}\\ 
\small{Supplementary Information is available in the online version of the paper.}\\

\noindent {\bf Competing Interests}\\ 
\small{The authors declare no competing financial interests.}\\

\noindent {\bf Reprints}\\ 
\small{Reprints and permissions information is available online at www.nature.com/reprints.}\\

\noindent {\bf Correspondence}\\ 
\small{Correspondence and requests for materials should be addressed to M.C.B.~(email: mariachiara.braidotti@isc.cnr.it)}\\

\noindent {\bf Data Availability}\\ 
\small{Data supporting the reported results and other findings of this study are available from the corresponding author upon reasonable request.}

\end{document}